\documentclass[reprint,amsmath,amssymb,aps]{revtex4-2}

\usepackage{natbib}
\usepackage{graphicx}
\usepackage{dcolumn}
\usepackage{bm}

\usepackage{graphicx}	

\newcommand{\rmin}{r_\mathrm{min}}
\newcommand{\rmax}{r_\mathrm{max}}
\newcommand{\dmax}{d_\mathrm{max}}
\newcommand{\sigd}{\sigma_\mathrm{d}}

\newcommand{\muasyr}{\mathrm{\mu as/yr}}

\def\aap{A\&A}
\def\nat{Nature}              
\def\apj{ApJ}
\def\mnras{MNRAS}
\def\aj{AJ}

\begin{document}

\preprint{IPPP/19/14}

\title{ Proper Motions of the Satellites of M31}

\author{Ben Hodkinson}
\author{Jakub Scholtz}%
 \email{jakub.scholtz@durham.ac.uk}
\affiliation{%
 IPPP, Durham University, DH1 3LE, UK
}%

\date{\today}

\begin{abstract}
We predict the range of proper motions of 19 satellite galaxies of M31 that would rotationally stabilise the M31 plane of satellites consisting of 15-20 members as identified by Ibata et al. (2013). Our prediction is based purely on the current positions and line-of-sight velocities of these satellites and the assumption that the plane is not a transient feature. These predictions are therefore independent of the current debate about the formation history of this plane. We further comment on the feasibility of measuring these proper motions with future observations by the THEIA satellite mission as well as the currently planned observations by HST and JWST.
\end{abstract}

\maketitle

\section{Introduction}

\citet{2013Natur.493...62I} reported the existence of a planar sub-group of 15 satellites of the M31 galaxy. Of the 15 in-plane satellites, 13 are co-rotating, which suggests, but does not show, this plane is a kinematically stable structure. Moreover, the small width of the plane, roughly 13kpc, is a challenge to explain. As a result, it is a matter of active debate if this structure may be an indication of phenomena not predicted within standard $\Lambda$CDM cosmology \citep{Zentner:2005wh,Ibata:2014qoa,Cautun:2015dqa,Pawlowski:2018sys}.  In particular, \citet{2015ApJ...800...34G,2016MNRAS.460.4348B} have argued that kinematically stable planes are less common than accidental alignments in the standard $\Lambda$CDM scenario. Therefore, the question of whether this plane is a truly stable structure has important consequences for our understanding of the $\Lambda$CDM model.

A similar plane of satellites has previously been discovered for the Milky Way \citep{2005A&A...431..517K}, and the proper motions (PMs) of the satellites for which data was not already available were predicted under the assumption that the MW plane is rotationally stabilised \citep{2013MNRAS.435.2116P, 2014ApJ...790...74P,2015MNRAS.453.1047P,2018A&A...619A.103F}. 

In this paper we carry out a similar analysis for the satellite members of the plane of M31 based  on our current knowledge of the satellites' orbital parameters, which includes their full three-dimensional position and line-of-sight velocities  \citep{2012AJ....144....4M,2012ApJ...758...11C}. At first glance, a permanent membership of an individual satellite in the plane can be reduced to the condition that the satellite's angular momentum is aligned with the normal to the plane. This fixes one of the components of proper motion. The remaining component determines the size of the angular momentum of the satellite and is limited by the condition that the satellite is on a bound orbit. As a result the constraints implied by the stability of the plane result in a finite range of allowed values of their proper motions.

In this paper, we identify the range of possible PMs for the 13 co-rotating satellites, two counter-rotating satellites and five additional satellites that were also reported in \citet{2013Natur.493...62I} as being consistent with a stable plane. Our results are a prediction and provide a benchmark against which measurements of the PMs can be compared to determine whether the plane is indeed a stable structure or a temporary alignment of satellites of M31.

We do feel the need to re-iterate two points: First, the results of this work are independent of the debate about the nature of this plane: it could be a rare structure of the $\Lambda$CDM model or a manifestation of a deviation from the $\Lambda$CDM. As long as the M31 plane is a stable feature, our results remain valid. Second, whether the plane is kinematically stabilised or an accidental alignment only observable in the current era is not a dichotomy: the answer may be that a subset of satellites form a kinematically stable plane, while the rest of the satellites are aligned accidentally.

This paper is organised as follows: in sec.~\ref{sec:method} we discuss our methods for determining properties of the satellites' orbits. In sec.~\ref{constraints} we define the constraints an orbit needs to satisfy in order to belong to the plane. In sec.~\ref{sec:results} we discuss our results and the experimental program needed to probe these results. We conclude in sec.~\ref{sec:conc}.

\section{Method}
\label{sec:method}

\subsection{Positions and Velocities of the M31 System}
We use the data available in \citet{2012AJ....144....4M} and in \citet{2012ApJ...758...11C} as presented in Table~\ref{data}. We take the velocity of M31 with respect to the Sun, according to \citet{2012AJ....144....4M} and \citet{2019ApJ...872...24V}, to be
\begin{align}
v_{\mathrm{los},\mathrm{M31}} &= -300\pm4\;\mathrm{km/s}\nonumber\\
\mu_{\alpha*,\mathrm{M31}} &= +65\pm 18\;\muasyr\\
\mu_{\delta,\mathrm{M31}} &= -57\pm 15\;\muasyr.\nonumber
\end{align}
In this paper, we will adopt a Cartesian coordinate system, aligned with the standard Galactic coordinate system, but centered on M31. 

\subsection{Properties of the Plane of M31 Satellites}

For our purposes, it is best to describe the plane by a unit normal vector $\hat{n}$ and the offset of the plane away from the origin (center of M31) along $\hat{n}$, $d$.  Although each individual satellite orbit is contained in a plane that also contains the origin (the satellites orbit around M31), it does not mean that the plane of satellites itself has to contain the origin unless all the orbits are exactly co-planar. 

We choose $\hat{n}$ and $d$ such that the sum of the squares of distances of the satellites from the plane is minimised. Note, that we exclude M32, NGC 205, LGS 3, IC10 and IC1613 in this calculation, as their planar membership is debatable, and consider just the first 15 satellites in Table~\ref{data}. Denoting the position vectors of satellites by  $\vec{r_i}$, then the sum of the squares of the distances is given by
\begin{equation}
D^2=\sum_{i=1}^{15} D_i^2 = \sum_{i=1}^{15} |\vec{r}_i\cdot\vec{n} + d|^2.
\end{equation}
We minimise $D$, subject to the normalisation constraint $|\vec{n}|=1$. This yields the normal vector and plane offset: 
\begin{align}
\hat{n} &=[0.887, 0.443,-0.132]\label{normvec}\\\nonumber
d &= - 3.5\pm1.5\mathrm{kpc},
\end{align}
where the uncertainty bands come from sampling the heliocentric distances of M31 and all the satellites from Gaussians with widths given by the uncertainty bands in Table~\ref{data}. We omit the uncertainty bands for the pole, because the pole of the plane $\hat{n}$ only varies by $1^\mathrm{o}$ (68\% confidence interval) under the same sampling.

\begin{table}
\centering
\begin{tabular}{lllll}
\hline
Name & R.A. & Dec. & D[kpc] & $v_{\text{los}}$[km/s] \\
\hline
And I & $00\;45\;39.8$ & $38\;02\;28$ & $727^{+18}_{-17}$ & $-376.3^{+2.2}_{-2.2}$ \\
And III & $00\;35\;33.8$ & $36\;29\;52$ & $723^{+18}_{-24}$ & $-344.3^{+1.7}_{-1.7}$ \\
And IX & $00\;52\;53.0$ & $43\;11\;45$ & $600^{+91}_{-23}$ & $-209.4^{+2.5}_{-2.5}$ \\
And XI & $00\;46\;20.0$ & $33\;48\;05$ & $763^{+29}_{-106}$ & $-419.6^{+4.4}_{-4.4}$ \\
And XII & $00\;47\;27.0$ & $34\;22\;29$ & $928^{+40}_{-136}$ & $-558.4^{+3.2}_{-3.2}$ \\
And XIII & $00\;51\;51.0$ & $33\;00\;16$ & $760^{+126}_{-154}$ & $-185.4^{+2.4}_{-2.4}$ \\
And XIV & $00\;51\;35.0$ & $29\;41\;49$ & $793^{+23}_{-179}$ & $-480.6^{+1.2}_{-1.2}$ \\
And XVI & $00\;59\;29.8$ & $32\;22\;36$ & $476^{+44}_{-29}$ & $-367.3^{+2.8}_{-2.8}$ \\
And XVII & $00\;37\;07.0$ & $44\;19\;20$ & $727^{+39}_{-25}$ & $-251.6^{+1.8}_{-2.0}$ \\
And XXV & $00\;30\;08.9$ & $46\;51\;07$ & $736^{+23}_{-69}$ & $-107.8^{+1.0}_{-1.0}$ \\
And XXVI & $00\;23\;45.6$ & $47\;54\;58$ & $754^{+218}_{-164}$ & $-261.6^{+3.0}_{-2.8}$ \\
And XXVII & $00\;37\;27.1$ & $45\;23\;13$ & $1255^{+42}_{-474}$ & $-539.6^{+4.7}_{-4.5}$ \\
And XXX & $00\;36\;34.9$ & $49\;38\;48$ & $681^{+32}_{-78}$ & $-139.8^{+6.0}_{-6.6}$ \\
NGC 147 & $00\;33\;12.1$ & $48\;30\;32$ & $712^{+21}_{-19}$ & $-193.1^{+0.8}_{-0.8}$ \\
NGC 185 & $00\;38\;58.0$ & $48\;20\;15$ & $620^{+19}_{-18}$ & $-203.8^{+1.1}_{-1.1}$
\\ \hline
M32 &  $00\;42\;41.8$ & $40\;51\;55$ & $805^{+82}_{-74}$ & $-199.0^{+6.0}_{-6.0}$ \\
NGC 205 &  $00\;40\;22.1$ & $41\;41\;07$ & $824^{+27}_{-26}$ & $-286.5^{+0.3}_{-0.3}$ \\
IC 10 &  $00\;20\;17.3$ & $59\;18\;14$ & $794^{+45}_{-43}$ & $-348.0^{+1.0}_{-1.0}$ \\
LGS 3 &  $01\;03\;55.0$ & $48\;20\;15$ & $769^{+25}_{-24}$ & $-203.8^{+1.1}_{-1.1}$ \\
IC 1613 &  $01\;04\;47.8$ & $02\;07\;04$ & $755^{+43}_{-41}$ & $-231.6^{+1.2}_{-1.2}$ \\
\hline
\end{tabular}
\caption{Data from \citep{2012AJ....144....4M,2012ApJ...758...11C} for the satellites reported in \cite{2013Natur.493...62I} to plausibly lie in the planar structure. The heliocentric distances are given in kpc, with upper and lower errors. The heliocentric velocities are given as line-of-sight velocities relative to the sun with upper and lower errors. The top section of this table contains presumed members of the M31 plane, while the lower section consists of satellites whose membership is debatable.}
\label{data}
\end{table}

Given $\hat{n}$ and $d$, we can also compute the thickness of the plane:
\begin{equation}
    \sigd = \frac{D}{\sqrt{15}} = 12.4\pm0.8\mathrm{kpc},
    \label{sd}
\end{equation}
where the uncertainty band comes from the same distance sampling as described above.

\subsection{Determining Satellite Orbits}

Each choice of proper motion $(\mu_{\alpha*},\mu_\delta)$ is equivalent to a different orbit. As we will see in section~\ref{constraints}, we are interested in three properties of each orbit: the maximum distance of the satellite from the plane in the next $10$ periapses
\begin{equation}
\dmax = \max_t |\vec{r}(t)\cdot\hat{n}+d|,
\end{equation}
and the minimum and maximum distances of the satellite from the center of M31 (also over the next 10 periapses)
\begin{align}
\rmin &= \min_t |\vec{r}(t)|\\
\rmax &= \max_t |\vec{r}(t)|.
\end{align}

Here we should briefly mention our choice of integration time. A possible choice would be to set a fixed integration time of order the age of M31, and we believe that this is an acceptable choice. However, there are orbits, which are only accidentally aligned with the plane and stay aligned for up to a couple billion years. 

An example of such an orbit would be a highly eccentric orbit such that the plane of this orbit is not aligned with the plane of M31, but the major axis of this orbit lies along the intersect of these two planes. Because orbits in the potential of an NFW halo are not closed, after the next periapsis the major axis of the orbit changes significantly and the orbit is no longer close to the plane. These orbits represent temporary alignments between the orbit and a plane unrelated to the orbit. 

Since we are interested in finding orbits that are not temporarily aligned with the plane of satellites, we would like to avoid accepting these orbits. In order to remove them, we choose to integrate over a fixed number of periapses: we chose this fixed number to be 10.

For comparison with other work we also compute the angle between the normal to the plane from Eq.~(\ref{normvec}) and the angular momentum of the satellite $\vec{L} = \vec{r} \times \vec{v}$ around the center of M31:
\begin{equation}
 \cos \theta = \vec{L}\cdot\hat{n}/|\vec{L}|.
 \label{ctheta}
\end{equation}

In order to compute these quantities we use GALPOT \citep{2016ascl.soft11006M} to integrate the orbits in the gravitational field of M31. We model M31 as re-scaled DM halo of the Milky Way. We use an NFW dark matter halo with density profile
\begin{equation}
\rho(r) = \frac{\rho_0}{(r/r_0)(1+r/r_0)^2},
\label{pot}
\end{equation}
where $\rho_0=1.7\times10^7$ M$_{\odot}$kpc$^{-3}$ (corresponds to $M_{\mathrm{M31}} \sim 2 M_{\mathrm{MW}}  \sim 2.6\times 10^{12} M_{\odot} $) and the scale radius $r_s=19$kpc  \citep{2017MNRAS.465...76M}. We have also produced results for a Milky Way-like potential with $M_{\mathrm{M31}} \sim M_{\mathrm{MW}}$.

We have omitted the M31 baryonic disk and bulge as well as the fact that its DM halo is elliptical. While the effects of the baryonic disk can be neglected, the effects of the non-spherical DM halo may be significant. Deviation from spherical symmetry of the background potential will contribute to changes of the angular momentum of each satellite. However, we leave this discussion to future work, where we would like to explore this effect in order to place a bound on the ellipticity of the M31 DM halo.

\subsection{Uncertainties}
\label{exp_uc}
We have found that varying the line-of-sight velocities within the uncertainty band in Table~\ref{data} does not significantly alter our results, so we do not include the impact of these variations in our work.

On the other hand, the uncertainties in the heliocentric distance measurements to the satellites are much more significant. To include this effect, we evaluate $\rmax$, $\rmin$ and $\dmax$ for five values of distance:
\begin{equation}
    \Delta \in \{\Delta_0+\epsilon_+,\Delta_0+\epsilon_+/2,\Delta_0,\Delta_0-\epsilon_-/2,\Delta_0-\epsilon_-\},
    \label{distset}
\end{equation}
 where $\Delta_0$ is the central value of the distance and $\epsilon_{+}$ and $\epsilon_-$ are the upper and lower edges of the one sigma uncertainty band from Table~\ref{data}.
 
We have also propagated the uncertainty due to the proper motion of M31. This offset can be treated as constant across the whole range of proper motion we show in our plots. In order to make this systematic uncertainty apparent, we show the M31 PM uncertainty in every plot in Figs.~\ref{theplot} and \ref{theotherplot} as a green ellipse.

\section{Constraints}
\label{constraints}
If the plane of satellites is a permanent feature, we expect it to be rotationally stabilised. This implies that the orbits of the in-plane satellites are constrained to be bound, the satellites must be able to survive on their orbits and they do not make large excursions away from the plane.

The first two criteria are not hard to implement: the orbit is bound if the energy of the satellite is negative
\begin{equation}
    E < 0.
\label{cr1}
\end{equation}

A slightly more stringent condition would be to require that the maximum excursion of the satellite away from M31 is less than the virial radius of M31 (500kpc). Otherwise, the satellite would not be identified as M31's satellite and would eventually cross into the Milky Way's gravitational potential. This criterion is then
\begin{equation}
    \rmax < 500\mathrm{kpc},
\label{cr1b}
\end{equation}
which replaces the previous criterion from Eq.~(\ref{cr1}). We would like to note that if a satellite is orbiting in the M31 plane of satellites, it is only a matter of time before its orbit's major axis is pointing at the MW, because orbits in the NFW profile precess fairly rapidly and the M31 plane is nearly edge-on to the MW. As a result, excursions past the virial radius are very likely to result in an eventual loss of the satellite.

If a satellite galaxy comes too close to M31 the tidal interactions strip its content and disrupt the satellite. Therefore, we impose a bound on the satellite's closest approach to M31. In this work we choose to implement this bound as:
\begin{equation}
    \rmin > 15\text{kpc}.
\label{cr2}
\end{equation}
We can think of this scale as an approximate Roche limit for a satellite of mass of order $10^7 M_\odot$ of size $0.5$kpc orbiting an enclosed mass of order $10^{11} M_\odot$. This is a conservative bound, because in reality such close encounters are still somewhat dangerous. 

The third criterion can be implemented in two different ways. We could require that in order to be a permanent member of the plane the angular momentum of the satellite has to be aligned with the normal to the plane.  This is equivalent to demanding that the $\theta$ from Eq.~(\ref{ctheta}) is smaller than some reference angle $\theta_0$. The angle $\theta_0$ is related to the scatter of these angles for satellites we consider in-plane. The authors of \citep{2013MNRAS.435.2116P, 2014ApJ...790...74P} have done just that and have chosen $\theta_0 = 37^\circ$ for the MW plane (the somewhat large size of this angle is driven by the fact that the MW plane is thicker than the M31 plane). We could determine $\theta_0$ from the scatter of the $\vec{n}$ on the unit sphere and this method will be discussed in the upcoming work \citep{futurework}. 

However, even if $\theta < \theta_0$, it is still possible for a satellite to stray far from the plane by a large radial excursion, which would make it appear as a non-member. For this reason we require that each satellite does not make excursions far away from the plane by requiring that:
\begin{equation}
    \dmax < \{1,2,3\}\times \sigd \approx \{13,25,37\}\text{kpc},
\label{cr3}
\end{equation}
where $\sigd$ is defined in Eq.~(\ref{sd}). Another advantage of expressing the third criterion this way is that $\sigd$ is a directly observable scale and can be determined from the data. 

A choice of $(\mu_{\alpha*},\mu_\delta)$ that satisfies the conditions in Eqs.~(\ref{cr1b}), (\ref{cr2}) and (\ref{cr3}) leads to an orbit that is consistent with a stable plane composed of the first 15 satellites in Table~\ref{data}.

\section{Results and Discussion}
\label{sec:results}
\subsection{Proper Motion Predictions}
\begin{figure*}
\begin{center}
\includegraphics[width =\textwidth]{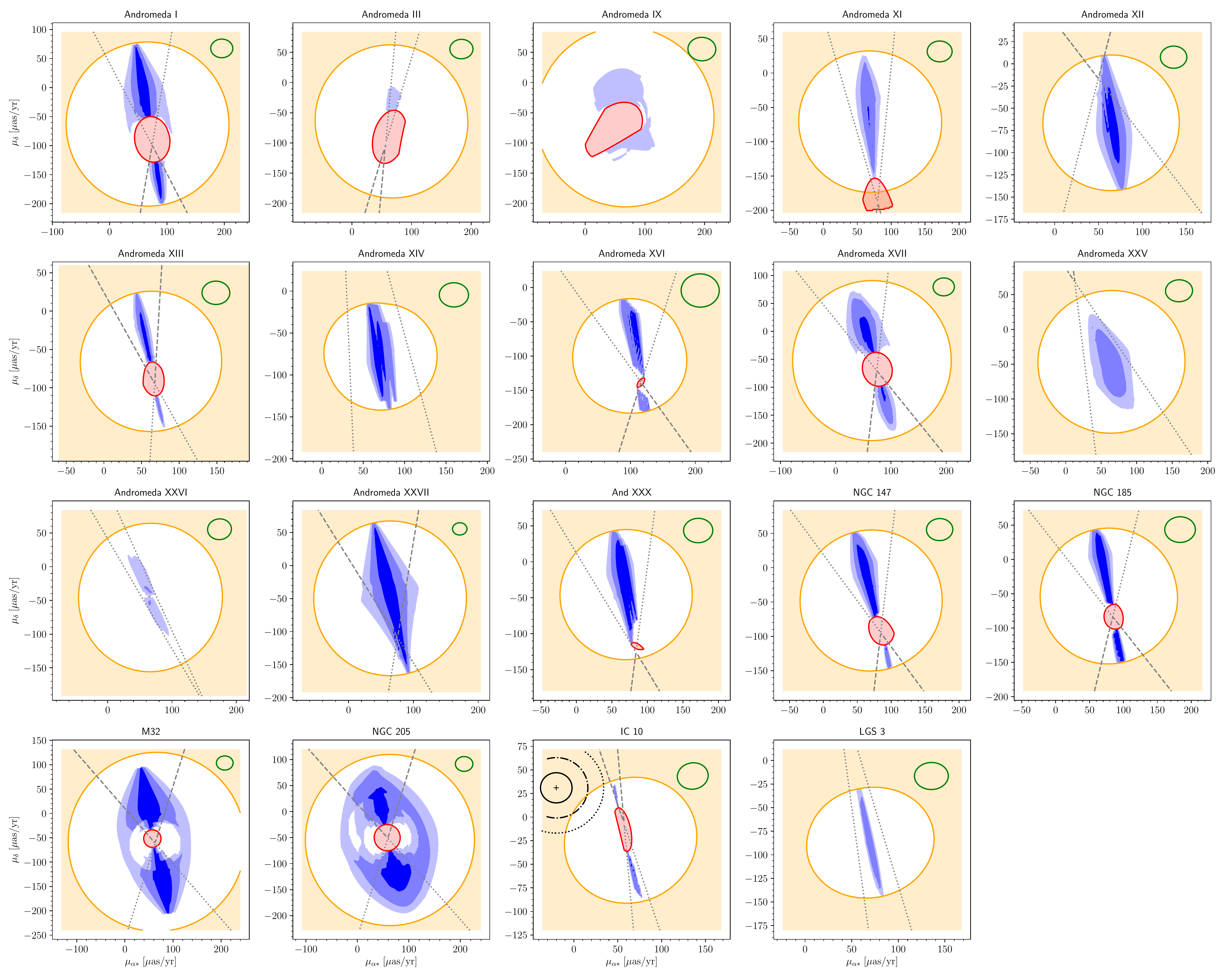}
\caption{Predicted PMs of satellites of M31. The blue region indicates PMs that are consistent with the satellite staying within $\{1,2,3\}\sigma_\mathrm{D}$ of the plane from darkest to lightest. The red region indicates PMs for which the satellite comes closer than 15kpc to the center of the M31 galaxy and is very likely to get disrupted.  The orange region indicates PMs for which the satellite wonders off more than 500kpc from the M31 galaxy. All of the previous regions are marginalised over one sigma band of distance measurements. The grey lines indicate the regions for which the angle between the satellite's angular momentum and the normal to the plane of satellites is less than 30 degrees (for the central value of satellite distance). The dotted lines indicate $\vec{L}$ and $\hat{n}$ are almost aligned, while the dashed lines enclose a region in which they are anti-aligned. Finally, the green circle indicates the uncertainty on the PM of the M31 galaxy itself. The top three rows (first 15 satellites) have been identified by \citet{2013Natur.493...62I} as very likely members of the plane. The bottom row (last five satellites) are possible members. The plot for IC1613 is missing because there are no proper motions that keep IC1613 within 500 kpc of M31. The plot of IC10 also shows the measurement of its proper motion by \citet{2007A&A...462..101B} and the $1\sigma$, $2\sigma$ and $3\sigma$ combined error-ellipsoids are shown as solid black, dash-dotted black and dotted black.\label{theplot}}
\end{center}
\end{figure*}

\begin{figure*}
\begin{center}
\includegraphics[width =\textwidth]{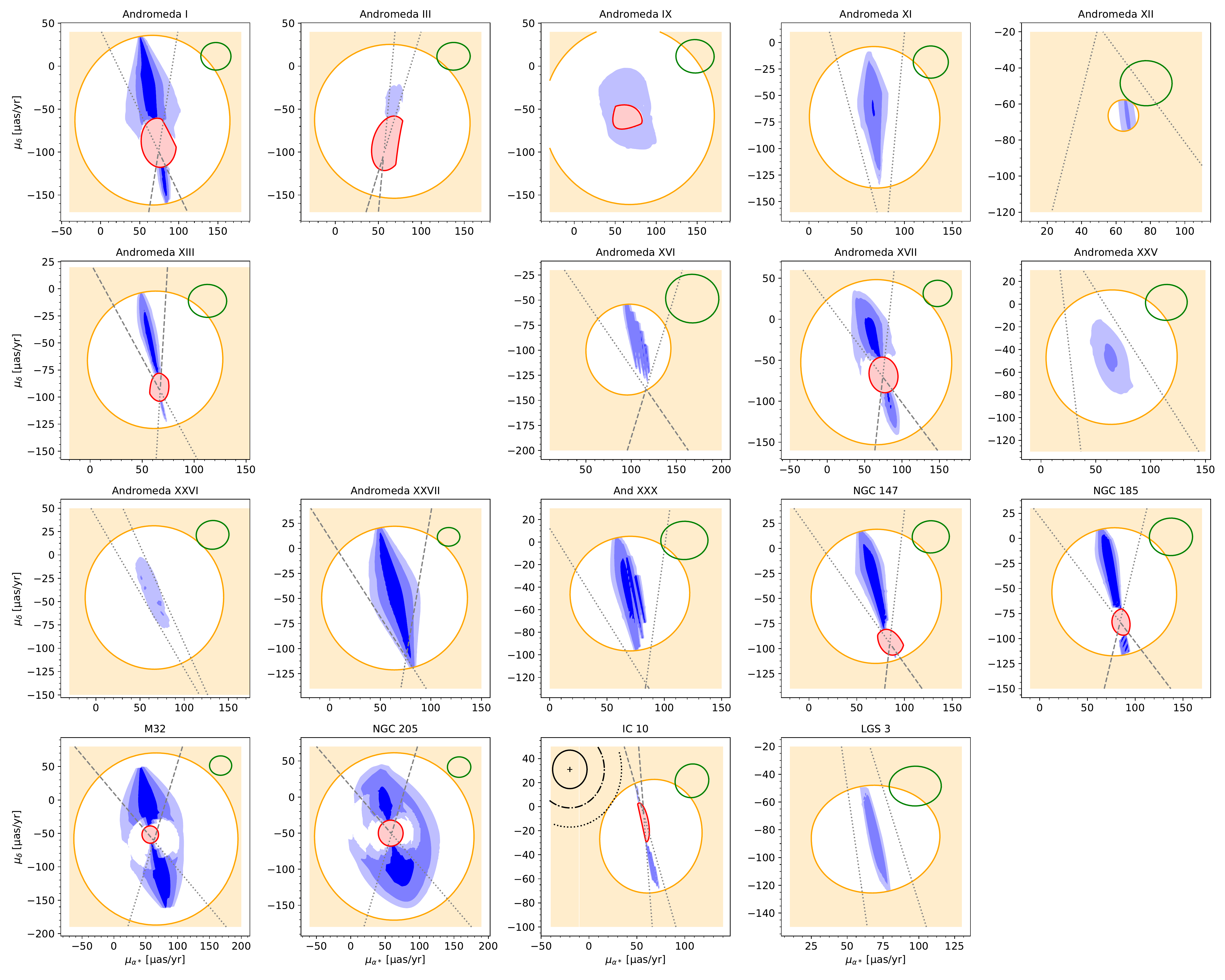}
\caption{Predicted proper motions of satellites of M31 with the ansatz $M_{\mathrm{M31}} = M_{\mathrm{MW}}$. The color scheme is identical to the one used in Fig.~\ref{theplot}. The plots for IC 1613 and Andromeda XIV are missing because there are no bound solutions for light $M_{\mathrm{M31}}$. \label{theotherplot}}
\end{center}
\end{figure*}

We present our results as a phase plot in $\mu_{\alpha*}$ and $\mu_\delta$  in Figs.~\ref{theplot} and~\ref{theotherplot}. The trajectory for each PM is tested against the constraints for five heliocentric distances from the set in Eq.~(\ref{distset}), as discussed in section \ref{exp_uc}.

The coloured regions are defined as follows:
\begin{enumerate}
\item The blue regions contain values of PMs that pass the constraints, and are consistent with the satellite staying within $\sigd$, $2\sigd$ and $3\sigd$ of the plane in at least one distance realisation (from darkest to lightest). 
\item The orange region contains trajectories which stray further than 500kpc from M31, i.e. which fail condition (\ref{cr1b}) for all five distance realisations. The boundaries of these regions lie on velocity contours, so the constraint essentially imposes a maximum velocity. 
\item The red region indicates trajectories which come closer than 15kpc to the centre of M31, i.e. that fail condition (\ref{cr2}) in all five distance realisations. As expected these regions lie around the singularity of the angle contours, corresponding to orbits with small angular momentum. 
\end{enumerate}
\par
The uncertainty in the PM of the M31 galaxy itself is indicated by a green circle in each plot, as discussed in sec.~\ref{exp_uc}.
\par
We include grey dashed lines which indicate the regions for which the angle, $\theta$, between the satellite's angular momentum and the normal to the plane of satellites is less than 30 degrees (for the central value of satellite distance). In \citet{2013MNRAS.435.2116P,2014ApJ...790...74P,2015MNRAS.453.1047P}, MW satellites with $\theta<37^\mathrm{o}$ were considered to be co-orbiting within the MW's plane of satellites. This approach produces predictions which are independent of the mass of M31, as a larger M31 mass allows higher velocity orbits along the same angular direction. However higher velocity orbits can deviate from the plane significantly for even a small value of $\theta$, and the $\theta<37^\mathrm{o}$ region on its own fails to include a restriction on such trajectories. For this reason, we believe Eq.~(\ref{cr3}), which defines our blue regions, gives a better indication of orbits which should truly be considered as stable members of the plane. \par
The blue regions for most of the satellites are centred on $\theta=0^\mathrm{o}$ or $180^\mathrm{o}$, indicating a direction of rotation. The directions of rotation are consistent with those suggested by \citet{2013Natur.493...62I}, where Andromeda XIII and Andromeda XXVII counter-rotate and the other 13 satellites co-rotate. Our results indicate it may also be possible for Andromeda I and XVII to counter-rotate within the plane, though the counter-rotating regions pass closer than 15kpc to the centre of M31 for at least one distance realisation. \par
We have also tested every other satellite of M31 (those not considered in-plane by \citet{2013Natur.493...62I}). None of the other satellites have PMs corresponding to orbits within the plane. \par
The plots in Figs.~\ref{theplot} and \ref{theotherplot} can be grouped into three broad categories with similar features:
\begin{enumerate}
    \item M32, NGC205 and Andromeda IX are M31's three closest satellites, with distances 23kpc, 42kpc and 40kpc from the centre of M31 respectively. As long as their velocity is small enough, there are orbits consistent with the plane for all angles between the satellite's angular momentum and the plane normal. It is likely these satellites are not truly a part of the planar structure, with their position only consistent with the plane due to their close proximity to the origin of M31. Indeed, \citet{2013Natur.493...62I} considered M32 and NGC205 as less likely planar members for this reason. The discontinuous nature of the contours in plots for M32 and NGC205 is caused by the fact that the distance uncertainties are significant compared to their actual distance from M31 and hence our five samples from the uncertainty band are not fine enough to produce smooth results. However, given most likely accidental membership in the plane, we feel it is unnecessary to determine the precise nature of these satellites' orbits.
    \item IC10, LGS3, IC1613 and Andromeda XVI are very far from M31, with distances 252kpc, 269kpc, 520kpc and 279kpc respectively. As a result, only a thin strip of PM's, which correspond to angular momenta very closely aligned to the plane normal, produce trajectories which remain closer than $3\sigma_\mathrm{D}$ from the plane. IC1613 is so far from the origin of M31 that no PM's correspond to planar orbits, so we don't include it in Figs.~\ref{theplot} and \ref{theotherplot}. Andromeda XVI on the other hand currently lies almost perfectly on the plane, so there is a relatively wide band of PM's consistent with it remaining less than $\sigma_\mathrm{D}$ from the plane. IC10, LGS 3 and IC1613 are more likely accidentally aligned with the plane.
    \item Andromeda III, XXV and XXVI are the furthest offset from the plane of those considered likely to be members by \citet{2013Natur.493...62I}. As a result the regions consistent with the plane are much smaller (in the Andromeda XXV plot there are no orbits within $1\sigd$ of the plane), with too large a velocity or too large an angular momentum deviation from the pane normal resulting in orbits which stray further than $3\sigma_\mathrm{D}$ from the plane.
    \item The remaining satellites were all considered likely members of the plane by \citet{2013Natur.493...62I}, and have similar spade-like blue regions in Figs.~\ref{theplot} and \ref{theotherplot}. This is due to a relationship between the variable $\dmax$ in constraint (\ref{cr3}) and $\cos \theta$ as defined in Eq.~(\ref{ctheta}). From the geometry of the setup it is obvious that $\dmax \leq \sin(\theta) \rmax$. However, since the orbits are not closed, over time this inequality is saturated. This explains the shape of the contours: instead of wedges of constant angle, we get shapes that narrow towards larger velocities because larger velocities imply larger $\rmax$ and hence to keep $\dmax$ constant, the acceptable range of $\theta$ must become smaller.
\end{enumerate}
 \par
 Finally, we would like to discuss the effect of mass of the host galaxy: the plots in Figs.~\ref{theplot}~and~\ref{theotherplot} represent the choices $M_{\mathrm{M31}} \sim 2M_{\mathrm{MW}}$ and $M_{\mathrm{M31}} \sim M_{\mathrm{MW}}$. The results are almost identical, up to a rescaling by a factor of $\sim\sqrt{2}$. This is expected because doubling the mass increases the escape velocities by a factor of $\sqrt{2}$. If the line-of-sight velocity is relatively small, then the escape velocity is saturated by the proper-motion which results in the size of the orange regions to scale as $M^{1/2}$ as we see. However, if the line-of-sight velocity component is close to saturating the escape velocity alone, then the proper motion components are much more influenced by the change of mass of M31. As an example the blue regions for Andromeda XII and XIV are more than a factor of $\sqrt{2}$ larger in Fig.~\ref{theplot} compared to Fig.~\ref{theotherplot}. This is because Andromeda XII and XIV have a large velocity relative to M31, at -272 and -208 km/s (for zero relative proper motion). A larger M31 mass therefore prevents these satellites from making large excursions from M31 and the plane. For the same reason, the red regions in Fig.~\ref{theplot} encroach further on the blue regions, proportionally, than in Fig.~\ref{theotherplot}. This is most noticeable for Andromeda IX and XI. 
 
 This brings about an interesting opportunity. When the proper motions of M31's satellites are measured, we can infer a bound on mass of M31 under the assumption that each satellite is a member of a stable plane of satellites.
 
 Similarly, once the proper motions are measured, it is possible to put a constraint on the proper motion of M31, by finding the M31 proper motion that minimises the $\dmax$ for all the measured satellites.
 
\par

\subsection{Proper Motion of IC10}

The existence of a maser in IC10 allowed the authors of \citet{2007A&A...462..101B} to measure the PM of this dwarf galaxy:
\begin{align*}
    \mu_{\alpha*,\mathrm{IC10}} =& -20\pm 5 \muasyr\\
    \mu_{\delta,\mathrm{IC10,exp}} =& +31\pm 8 \muasyr.
\end{align*}
We have shown the value of this measurement in both Figs.~\ref{theplot} and ~\ref{theotherplot}. The 1$\sigma$, 2$\sigma$ and 3$\sigma$ error-ellipsoids are the result of combining the error bars from the PM of M31 and PM of IC10 in quadrature. For heavier M31, Fig.~\ref{theplot}, the bound orbit is within 2$\sigma$ and technically still not ruled out. However, for light M31, all bound orbits with apsis less than 500kpc are ruled out by 3$\sigma$. In both scenarios, all in-plane orbits are ruled out by at least 3$\sigma$ and therefore it is safe to conclude the IC10 is unlikely to be a member of the plane.

\subsection{Experimental Feasibility}
Fig.~\ref{theplot} illustrates that in order to distinguish the random alignment hypothesis from the coherent structure hypothesis it is necessary to measure one of the linear combinations of the PM to accuracy of order $10$ $\muasyr$. Naturally we need to ask if this is feasible.

There are two planned measurements by HST and JWST. The HST has already observed the proper motions of NGC147 and NGC 185, with planned accuracy of order 25km/s, which corresponds to PMs of order $10\;\muasyr$ \citep{sohn}. The JWST has planned guaranteed time observation for the PMs of Andromeda I, III, XIV, and XVII \citep{vdMarel} with similar planned accuracy. As a result in a couple of years we should have information about six out of the fifteen members of the M31 plane.

Unfortunately, GAIA is unlikely to deliver in this direction. The horizontal branch (HB) in Andromeda I has magnitude around 25 \citep{Costa:1996za} in the V band.
This is  similar for most of the M31 satellites as the distance modulus does not vary too much (it is between 23.9 and 24.9 for all the M31 satellites). In the G band this corresponds to magnitude 25.5 for the center of Horizontal Branch stars.

The GAIA satellite has not been directly tested to measure PMs of such dim stars, but naively extrapolating the PM uncertainty relation from \citet{2016A&A...595A...1G} we arrive at an expected measurement uncertainty of order $10^5 \muasyr$. We would need to observe close to $10^8$ HB stars in order to determine the PM of Andromeda I to the desired accuracy. Since Andromeda I does not have this many HB stars (or this many stars for that matter), the situation seems hopeless even without dealing with additional caveats associated with our naive extrapolation.

However, the proposed THEIA mission \citep{2017arXiv170701348T} has better prospects. With 40 hours of observation per satellite over four years, THEIA should be able to achieve PM measurement with uncertainty of order $200\;\muasyr$ for stars of the 25th magnitude. As a result only a handful of stars (roughly 100) per satellite galaxy would be needed to start resolving the question of stability of the M31 plane. Given that even JWST is likely to fly before THEIA is approved, THEIA would only need to observe the remaining 9 in-plane satellites. As a result we would advocate that the THEIA mission would be able to resolve whether or not the M31 plane is a kinematically stable feature with a dedicated total of $\mathcal{O}(400)$ hours of observation.

\section{Conclusion}
\label{sec:conc}
We have collected the available data on the three-dimensional position and line-of-sight velocity for 20 of the satellites of M31, 15 of which are believed to form a narrow planar structure. We have calculated the range of possible PMs for each satellite consistent with this plane being a stable structure and the satellite being a member of the structure, by requiring the corresponding orbits to lie close to the plane. The PMs are consistent with the direction of rotation reported in \citet{2013Natur.493...62I}. Our results consist of predictions of PMs of these satellites and provide a benchmark for future astronomical measurements. Some of these proper motions will be published soon (NGC147 and NGC185) and others have been given guaranteed observation time by JWST (Andromeda I, III, XIV, and XVII). On top, the proposed THEIA mission \citep{2017arXiv170701348T} would be capable of delivering the resolution required to determine whether the true PMs are consistent with the sets presented here. This would help resolve the question of whether the plane is rotationally stabilised or a temporary structure. We have compared the previous measurement of the proper motion of IC10 and we conclude it is highly unlikely to be a proper member of the M31 plane of satellites.

Furthermore, if the measured PMs indicate the plane is at least partially stabilised, we could use the PMs to derive a bound on the mass of M31 and form an independent estimate of the PM of M31. 

\section*{Acknowledgements}
We would like to thank Jan Scholtz and Carlos Frenk for fruitful discussions. We would like to thank Marcel Pawlowski for many insights that made this paper much better. We would also like to thank Roeland van der Marel for his help. We would like to thank Steven Abel for support for this project as BH's summer advisor. JS would like to thank the COFUND scheme and the IPPP for his funding. BH would also like to thank the IPPP summer fellowship program for partial support. We have also benefited from the availability of the Astropy software package \citep{astropy:2013,astropy:2018}.


\bibliography{bibliography}

\end{document}